\documentclass[preprint]{article}
\usepackage{graphicx}
\usepackage{dcolumn}
\usepackage{bm}
\usepackage{amsmath}
\usepackage{amssymb}
\usepackage{multirow}
\usepackage{caption}
\usepackage{epstopdf}
\usepackage{hyperref}
\usepackage{cite}
\usepackage{subcaption}

\begin{document}
{\setlength{\oddsidemargin}{1.2in}
\setlength{\evensidemargin}{1.2in} } \baselineskip 0.55cm
\begin{center}
{\LARGE {\bf Bose-Einstein condensate stars in massive gravity}}
\end{center}
\date{\today}
\begin{center}
  Meghanil Sinha, Bharat Singh, S. Surendra Singh* \\
Department of Mathematics, National Institute of Technology Manipur,\\
Imphal-795004,India\\
Email:{ meghanil1729@gmail.com, bharatsinghiam@gmail.com, ssuren.mu@gmail.com}\\
 \end{center}
 
\textbf{Abstract}: This study explores the construction, validity and the properties of Boson or Bose-Einstein condensate (BEC) stars under the framework of de Rham-Gabadadze-Tolley (dRGT) like massive gravity, employing the Kuchowicz metric potential to model their internal structure. This gravitational framework accounts for a massive graviton while ensuring the absence of ghost instabilities during propagation. The BEC stellar configuration in this study was obtained by determining the solutions characterized by static and spherical symmetric metric. This study provides a detailed account of the stellar structure, highlighting the roles played by massive gravity and the Kuchowicz metric through a combination of analytical and numerical solutions. Our work specifically utilizes the Colpi-Wasserman-shapiro (CWS) and Gross-Pitaevskii (GP) equations of state (EoS) to model the internal thermodynamic behavior of the BEC. We have evaluated the physical viability of the BEC stellar framework by analyzing the energy conditions, and the EoS parameter along with the gradients of the energy-momentum tensor. The stability criteria such as the study of surface redshift, adiabatic index and squared sound velocity were utilized to confirm that our proposed model is both stable and physically consistent. Hence, this study offers a definitive structural analysis of the BEC stars, providing precise results in this massive gravity environment.\\


\section{Introduction}\label{sec1}

\hspace{1cm}BEC stars are particularly compelling because their equilibrium is maintained by quantum self-interactions making them distinct from other compact stars \cite{B1}. The exploration of such objects began with the study of geons-where stable equilibrium was achieved through the self-gravitating photons \cite{B2}. Subsequently, a distinct class of stellar objects was identified by deriving self-consistent solutions to the coupled Einstein-Klein-Gordon equations, forming the theoretical basis for the modern models \cite{B3}. In the period following these discoveries, the majority of the research was concentrated on the configurations of self-gravitating scalar fields \cite{B1,B4,B5}. The solutions derived in \cite{B3} characterized the fundamental ground state of these stellar objects, where the self-interaction terms of the scalar field was not taken into account. Subsequent research in \cite{B6} demonstrated that the introduction of even a marginal self-interaction term can profoundly modify the ground state solutions previously derived in \cite{B3,B7}. By incorporating an interaction term, the model accounts for an extra pressure component that reinforces the quantum dispersion pressure, effectively balancing the object's self-gravity. This leads to a stellar configuration with a significantly higher mass than the previously documentations in \cite{B3,B7}. Furthermore, the dynamical stability of these objects was subjected to a rigorous assessment through the application of perturbative methods \cite{B8,B9,B10}. The intriguing nature of BEC stars makes them excellent candidates for probing the limits of current physical models and testing the validity of diverse theoretical frameworks. Recent advancements in effective field theory have enabled the investigation of BEC stars that utilize vector meson mediation to provide necessary repulsive self-interaction \cite{B11}. A further possibility involves the existence of such objects composed of composite scalar fields. A composite scalar field can be generated when super conducting quark matter undergoes a BCS-BEC transition, shifting the state of matter from a BCS phase to a BEC phase \cite{B12,B13,B14,B15,B16,B17,B18,B19,B20,B21,B22,B23}.\\
In the limit of extremely low temperatures, a dilute Bose gas forms a condensate by populating the collective ground state, resulting in a characteristic sharp peak that emerges within the broader momentum and co-ordinate space distributions. The experimental realization of quantum degeneracy, made possible by advanced cooling techniques has paved the way for several research arenas in many body physics and gravitation \cite{GR1,GR2,GR3,GR4,GR5,GR6}. The accumulation of a vast number of particles within the identical quantum state is similar to the emergence of macroscopic quantum coherence, characterized by the collective occupation of a single quantum state. The formation of a BEC is a phase-transition phenomenon triggered when the thermal wavelengths of the constituent particles exceed their mean inter particle spacing, leading to a collective quantum identity. Achieving the condensate state requires high degree of inter particle correlation, which facilitates the transition from a disordered gas to a coherent many body system \cite{GR1,GR2,GR3,GR4,GR5,GR6}. Experimental advances have showed that the transition from BCS superfluidity to a BEC state represents a continuous evolution of fermion pairing, rather than distinct physical processes \cite{S1,S2,S3}. Self-gravitating BECs may account for a large portion of the Universe's dark matter (DM), offering an unique framework for understanding galactic structures. Extensive academic effort has been dedicated to investigating the behavior and implications of BECs in both cosmological and astrophysical backdrops \cite{MS,RR,S,GR}. Building upon the insights provided by the previous research, this study investigates the properties of self-gravitating BEC stars within the theoretical framework of dRGT massive gravity.\\
Current cosmological evidence fundamentally establishes that the Universe is presently in a state of accelerated expansion. The late-time acceleration of the cosmic expansion has been verified by a diverse array of independent observational evidences \cite{d1,d2,d3,d4,d5,d6,d7,d8}. Dark energy (DE), defined as a hidden energy source with negative pressure was first suggested to explain why the Universe's expansion is speeding up. In modern cosmology, $ \Lambda $ serves as one of the representatives of DE. By integrating cold DM into the standard cosmological model, a reliable model ($\Lambda$CDM) has been created that accounts for the Universe's current expansion and the distribution of the galaxies. Despite its successes, the model faces a lot of problems related to a massive discrepancy between the predicted and the observed values of the cosmological constant. Secondly, there is no clear reason why DE and no other matter should exist in their current proportions. Given its extensive empirical validation, general relativity (GR) stands as one of the most reliable gravitational framework for mathematically modeling and interpreting diverse cosmological phenomena. The success of the GR framework is best illustrated by the LIGO-VIRGO detection of gravitational waves \cite{d9}. The transition towards new gravitational theories is driven by the limitations of the standard GR model, where it fails to explain two critical phenomena : the rapid inflation of the Universe and the current accelerated expansion driven by DE \cite{d10}.\\
The mathematical foundation for the massive gravity was established in 1939 \cite{dr0}. This model was the first to describe a spin-2 field (gravity) by introducing a non-zero mass term for the graviton within a flat spacetime framework. The initial success of massive gravity was tempered by subsequent theoretical hurdles with time, among which it includes van Dam-Veltman-Zakharov (vDVZ) discontinuity, Boulware-Deser ghost and the Vainshtein mechanism \cite{dr1,dr2,dr3,dr4,dr5}. After several years, the 2010 formulation of the dRGT theory successfully accounted for the above problems \cite{dr6,dr7}. dRGT massive gravity represents a consistent extension of GR. By endowing the spin-2 graviton with a non-zero mass, this framework provides a rich landscape for theoretical research. This framework functions as a non-linear completion of the Einstein-Hilbert action. By incorporating the graviton with mass, this theory introduces cosmological solutions where the graviton's own mass term effectively mimics the repulsive pressure of a cosmological constant, potentially eliminating the need for a separate DE sector. Observational data for the rotation curves of the Milky Way and the LSB galaxies can be successfully modeled using the dRGT massive gravity as a stand-alone framework \cite{dr8}.\\
The literature provides various solutions for compact stellar remnants within the dRGT framework \cite{dr9,dr10,d,dr}. The discovery of charged black hole solutions in dRGT and bigravity  scenarios represents a significant theoretical advancement \cite{dr11,dr12}. While the massive gravity was once mostly a mathematical curiosity, the era of gravitational-wave astronomy has turned it into a testable reality with the data from LIGO and VIRGO regarding the merger of massive stellar remnants \cite{dr13,dr14}. Against the historical evolution of massive gravity, we have explored the BEC stellar model governed by the laws of dRGT gravity. Obtaining exact analytical solutions for the field equations in dRGT massive gravity model will present significant mathematical hurdles. Adopting a specific ansatz for the metric function serves as an essential analytical tool in this context. We have employed a predetermined metric function to facilitate a clearer analysis of stellar equilibrium and in order to understand the confinement of matter within the stellar boundary. The Kuchowicz metric serves as an optimal choice here for defining the interior spacetime geometry. By utilizing this specific metric, we can effectively regulate and describe the distribution of matter within the stellar framework, ensuring that the model remains physically consistent. This metric is characterized by its analytical regularity, remaining free of singularities making it an ideal candidate to study compact stellar objects.\\
Thus, in this study we have successfully modeled a BEC star characterized by the Kuchowicz metric under the dRGT massive gravity regime. The manuscript is presented as follows: Section (\ref{sec1}) provides an overall view of the dRGT massive gravity framework alongside the theoretical construct of BEC stars. Section (\ref{sec2}) lays out the mathematical core for the massive gravity and introduces the Kuchowicz metric in this context. Section (\ref{sec3}) investigates the thermodynamic properties of the BEC EoSs and its subsequent impact on stellar equilibrium. Subsequent analysis includes the energy conditions and an evaluation of the radial gradients of the stress-energy tensor components to ensure physical consistency. Section (\ref{sec4}) looks at the star's stability through different approaches and finally section (\ref{sec5}) provides our concluding remarks.\\

\section{Massive gravity formulation}\label{sec2}

\hspace{1cm}The action in dRGT massive gravity is given by\\
\begin{equation}\label{1}
\emph{S} = \frac{1}{16 \pi G}\int d^{4}x \sqrt{-g} \big[ R + M^{2}\sum_{i}^{4} c_{i}V_{i}(g,\Psi) \big] + L_{m}
\end{equation}\\
where $ L_{m} $ = matter Lagrangian, $ G $ = universal gravitational constant, $ g $ = determinant of the dynamical metric tensor $ g_{\alpha \beta} $, $ R $ = Ricci scalar curvature. Here $ M $ stands for the graviton's mass. We have taken into consideration $ G = c = 1 $ in geometrized units. The introduction of a non-vanishing graviton mass necessitates a modified effective potential within the four-dimensional spacetime manifold as\\
\begin{equation}\label{2}
V_{1} = [K]
\end{equation}
\begin{equation}\label{3}
V_{2} = [K]^{2} - [K^{2}]
\end{equation}
\begin{equation}\label{4}
V_{3} = [K]^{3} - 3[K][K^{2}] + 2[K^{3}]
\end{equation}
\begin{equation}\label{5}
V_{4} = [K]^{4} - 6[K^{2}][K]^{2} + 8[K^{3}][K] + 3[K^{2}]^{2} - 6[K^{4}]^{4}
\end{equation}\\
where the symbol $ K $ denotes the trace operation applied to the matrix $ K^{\alpha}_{\beta} = \sqrt{g^{\alpha\gamma}\Psi_{\gamma\beta}} $. By varying the action $ \emph{S} $ presented in equation (\ref{1}) w.r.t $ g_{\alpha\beta} $, we get\\
\begin{equation}\label{6}
R_{\alpha\beta} - \frac{1}{2}Rg_{\alpha\beta} + M^{2}\eta_{\alpha\beta} = 8 \pi T_{\alpha\beta} 
\end{equation}\\
where $ \eta_{\alpha\beta} $ is represented by
\begin{eqnarray}
\eta_{\alpha\beta} &=& -\frac{c_{1}}{2}(V_{1}g_{\alpha\beta} - K_{\alpha\beta}) - \frac{c_{2}}{2}(V_{2}g_{\alpha\beta} - 2V_{1}K_{\alpha\beta} + 2K^{2}_{\alpha\beta}
\nonumber \\
&& -\frac{c_{3}}{2}(V_{3}g_{\alpha\beta} - 3V_{2}K_{\alpha\beta} + 6 V_{1}K^{2}_{\alpha\beta} - 6 K^{3}_{\alpha\beta})
\nonumber \\
&& -\frac{c_{4}}{2}(V_{4}g_{\alpha\beta} - 4V_{3}K_{\alpha\beta}  + 12 V_{2}K^{2}_{\alpha\beta} - 24 V_{1} K^{3}_{\alpha\beta} + 24 K^{4}_{\alpha\beta}).
\end{eqnarray}\\
To model the BEC star, we use the following mathematical description for the spacetime structure as\\
\begin{equation}\label{8}
ds^{2} = - e^{2x}dt^{2} + e^{2y}dr^{2} + r^{2}(d \theta^{2} + \sin^{2}\theta d\phi^{2})
\end{equation}
where $ x $ and $ y $ depend on the radial co-ordinate. We will now use the reference metric consistent with the previous models while broadening the theory \cite{dr15} in
\begin{equation}\label{9}
\Psi_{\alpha\beta} = diag(0, 0, A^{2}, A^{2}\sin^{2}\theta)
\end{equation}\\
where $ A $ is defined as a positive valued parameter. By applying this reference metric specified in equation (\ref{9}), we get
\begin{equation}\label{10}
V_{1} = \frac{2A}{r} \hspace{0.5cm} V_{2} = \frac{2A^{2}}{r^{2}} \hspace{0.5cm} V_{3} = 0 \hspace{0.5cm} V_{4} = 0.
\end{equation}\\
Here, the stellar interior is modeled as a perfect fluid distribution, for which we have the energy-momentum tensor given as\\
\begin{equation}\label{11}
T_{\alpha\beta} = (\rho + p)u_{\alpha}u_{\beta} - pg_{\alpha\beta}
\end{equation}\\
where $ u_{\alpha} $ represents the four-velocity vector of the fluid, while $ \rho $ and $ p $ characterize the energy density and isotropic pressure distribution respectively. As a result, we have the non-vanishing components of the field equations as\\
\begin{equation}\label{12}
e^{-2y}(2ry' - 1) + 1 - M^{2}A(c_{2}A + c_{1}r) = 8 \pi r^{2}\rho
\end{equation}
\begin{equation}\label{13}
e^{-2y}(2rx' + 1) - 1 -M^{2}A(c_{2}A + c_{1}r) = 8 \pi r^{2} p
\end{equation}
\begin{equation}\label{14}
e^{-2y}(x' + rx'^{2} - y' - rx'y' + rx'') - \frac{M^{2}c_{1}A}{2} = 8 \pi r^{2} p.
\end{equation}\\
Here, first and second-order derivatives w.r.t to radial co-ordinate are represented using prime and double-prime notations. Thus, we have the conservation equation as\\
\begin{equation}\label{15}
p' + (\rho + p)x' = 0.
\end{equation}\\
Our interior solution is designed to match smoothly with the exterior Schwarzschild metric. The exterior spacetime is represented by\\
\begin{equation}\label{16}
ds^{2} = -(1 - \frac{2 \bar{M}}{r})dt^{2} + \frac{1}{(1 - \frac{2 \bar{M}}{r})}dr^{2} + r^{2}(d \theta^{2} + \sin^{2}\theta d\phi^{2})
\end{equation}\\
where $ \bar{M} $ = total mass. At $ r = R $, the boundary conditions give
\begin{equation}\label{17}
e^{2x(R)} = e^{-2y(R)} = 1 - \frac{2 \bar{M}}{r}.
\end{equation}\\
Now, with the help of the Kuchowicz metric formulation, we proceed to investigate the physical implications and the structural consequences within the dRGT gravitational framework. As a particular case of interest, we have assumed the metric $ e^{2y(r)} $ to follow the Kuchowicz profile, which ensures a regular and well-behaved geometry at the stellar center \cite{K,K1,K2}. It is given as
\begin{equation}\label{18}
e^{2y(r)} = e^{Er^{2} + 2\log[F]}
\end{equation}\\
where the parameters $ E $ and $ F $ serve as constants within the metric. $ E $ possesses the dimension of inverse length squared $[L^{-2}]$, whereas $ F $ is a dimensionless quantity. The model is mathematically sound and contains no breakdowns or singularities. Applying the boundary conditions at the stellar surface $(r=R)$ requires a smooth matching of the interior and exterior manifolds. Enforcing the continuity of $ g_{rr} $ and its radial derivative provide\\
\begin{equation}\label{19}
\frac{1}{1 - \frac{2 \bar{M}}{R}} = e^{ER^{2} + 2\log[F]}
\end{equation}
\begin{equation}\label{20}
-\frac{2 \bar{M}}{R^{2}(1 - \frac{2 \bar{M}}{R})^{2}} = (2ER)e^{ER^{2} + 2\log[F]}.
\end{equation}\\
The above constraints facilitate the determination of the dimensionless parameter which is presented below as\\
\begin{table}[h!]
\centering
\caption{ BEC stars modeled with Kuchowicz function in massive gravity }
\begin{tabular}{||p{4cm}|p{3.5cm}|p{3cm}|p{2.5cm}|p{2.0cm}||}
\hline\hline
\hspace{1cm} Star  & \hspace{0.5cm}$ Mass(\bar{M})(M_{\bigodot}) $ & \hspace{0.5cm} $ Radius (km) $ & \hspace{0.5cm} $E(km^{-2})$ & \hspace{0.5cm} $F$  \\
\hline\hline
$\hspace{1cm} PSR JI6142230 $ & \hspace{0.5cm}$ 1.97 $ & $ 10.3 $  &  $ 0.00292 $ &  $ 1.0899 $\\[1pt]
\hline
$\hspace{1cm} Vela X-1 $ & \hspace{0.5cm}$ 1.77 $ & $ 9.99 $  &  $ 0.00275 $ &  $ 1.0849 $ \\[1pt]
\hline
$\hspace{1cm} PSR J1903+327 $ & \hspace{0.5cm}$ 1.667 $ & $ 9.82 $  &   $ 0.00267 $ &  $ 1.0826 $ \\[1pt]
\hline 
$\hspace{1cm} Cen X-3 $ & \hspace{0.5cm}$ 1.49 $ & $ 9.51 $  &   $ 0.002523 $ &  $ 1.07668 $ \\[1pt]
\hline
$\hspace{1cm} SMC X-1 $ & \hspace{0.5cm}$ 1.29 $ & $ 9.13 $  &   $ 0.0023 $ &  $ 1.07003 $ \\[1pt]
\hline\hline
\end{tabular}
\label{Tab:4}
\end{table}\\

\section{BEC stellar properties through Colpi-Wasserman-Shapiro (CWS) and Gross-Pitaevskii(GP) EoS}\label{sec3}

\hspace{1cm}BEC stars are compact astrophysical bodies composed of BECs that are held together by their own gravitational pull \cite{GR,RR1}. The characteristics of BEC stars have been studied across a variety of physical contexts and theoretical frameworks. BEC stars have also been put forward as viable candidates for DM \cite{RR2}. The scope of BEC stellar models has been expanded to include the effects of finite temperature \cite{RR3}. To analyze the BEC stars under modified gravity theories, we utilize the following formalisms to characterize the underlying BEC matter. Here, we have adopted the relativistic CWS EoS derived from the scalar field theory given by \cite{B6}\\
\begin{equation}\label{21}
p(\rho) = \frac{1}{36N}(\sqrt{1 + 12N\rho} - 1)^{2}
\end{equation}\\
where $ N = \frac{\mu \hbar^{3}}{4m^{4}} $.  The parameter $ m $ here represents the mass of the condensate particle with $ \mu $ as a dimensionless scaling factor given as\\
\begin{equation}\label{22}
\mu = (9.523 \times 8\pi)(\frac{B}{1 fm})(\frac{m}{2 \grave{m}})
\end{equation}\\
where $ \grave{m} $ = nucleon mass and $ B $ = scattering length. Another popular way to describe a BEC star's internal matter is the GP EoS, which is defined as\\
\begin{equation}\label{23}
p(\rho) =  \frac{H \rho^{2}}{2 m^{2}}
\end{equation}\\
where $ H = \frac{4 \pi B \hbar^{2}}{m} $ is the interaction strength. The aforementioned BEC frameworks have been employed to investigate the configurations of static and rotating stellar models \cite{GR,RR1,RR2,RR3,RR4,RR5}. By assigning the condensate particle a mass of $ m = 2 \grave{m} $, with $ \grave{m} = 1.675 \times 10^{-24} g $, we model the system under the assumption that nucleons form composite bosons. This effectively treats the matter as a collection of Cooper-like pairs, allowing the fermionic nucleons to follow Bosonic statistics. In our study, we set the scattering length as $ B = 1 fm $, and based on this the interaction strength becomes $ H = 4.17 \times 10^{-43} g$ $ cm^{5} / s^{2} $. The numerical solution of the field equations (\ref{12}-\ref{14}) subject to the junction conditions yields the radial behavior of the metric component $ e^{2x(r)} $ where we have taken $ M^{2}c_{1} = 0.25 $, $ M^{2}c_{2} = 0.3 $ and $ A = 0.5 $. The sensitivity of these solutions to the varying physical parameters is depicted graphically in figures (\ref{1}) and (\ref{2}). The results demonstrate that the metric potential conforms to the theoretical expectations, exhibiting physically consistent and non-singular behavior throughout.\\

\begin{figure}[htbp]
    \centering
    \begin{minipage}{0.45\textwidth}
        \centering
        \includegraphics[width=\linewidth]{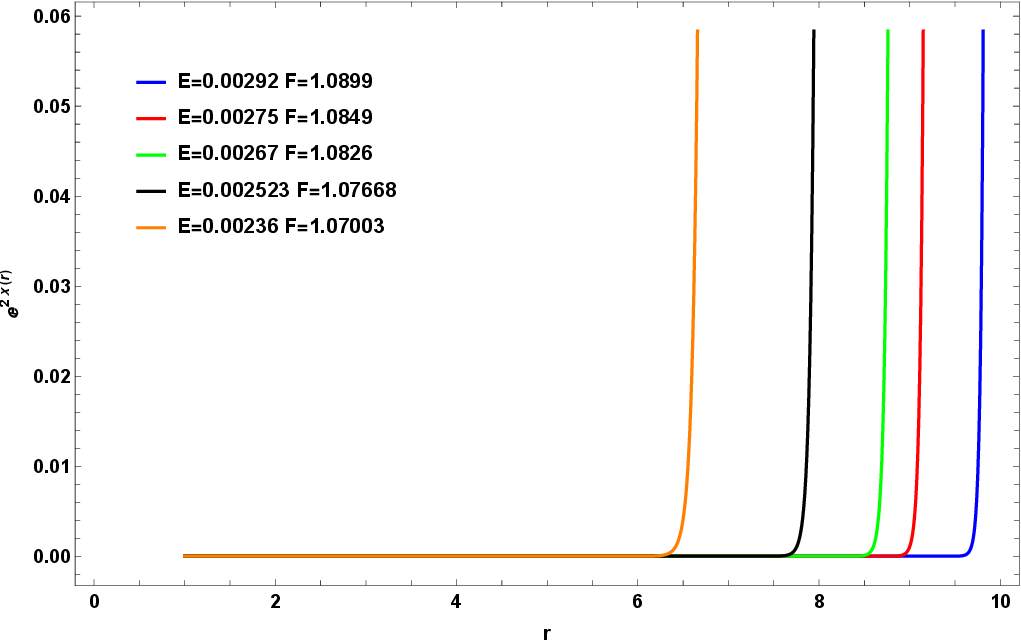}
        \caption{Plot of $ e^{2x} $ w.r.t radial co-ordinate for CWS EoS for $ E = 0.00292, F = 1.0899; E = 0.00275, F = 1.0849, E = 0.00267, F = 1.0826, E = 0.002523, F = 1.07668, E = 0.0023, F = 1.07003 $.}
        \label{1}
    \end{minipage}
    \hfill
    \begin{minipage}{0.45\textwidth}
        \centering
        \includegraphics[width=\linewidth]{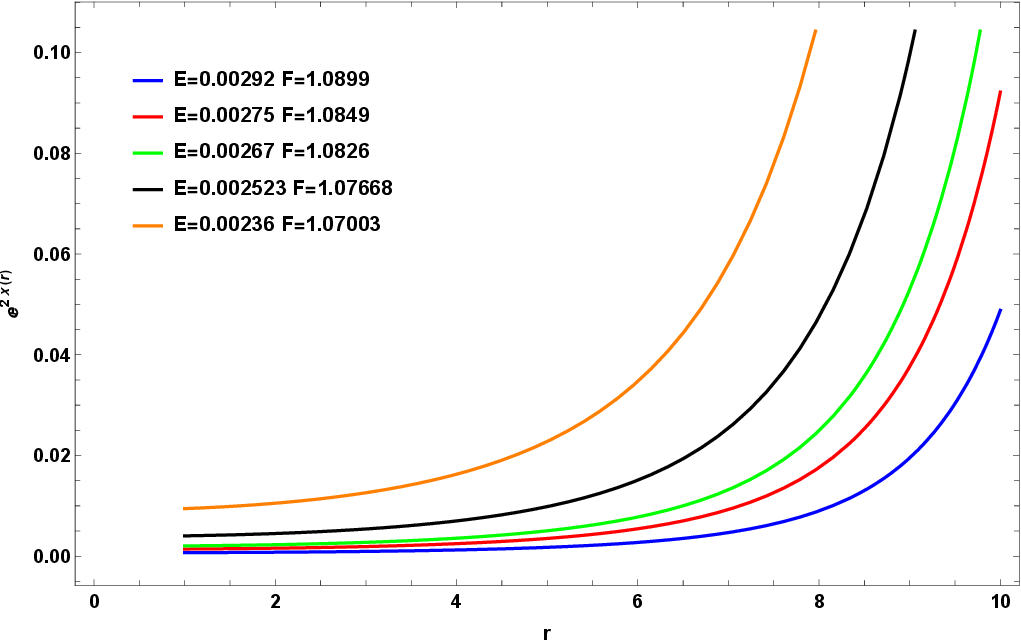}
        \caption{Plot of $ e^{2x} $ w.r.t radial co-ordinate for GP EoS for $ E = 0.00292, F = 1.0899; E = 0.00275, F = 1.0849, E = 0.00267, F = 1.0826, E = 0.002523, F = 1.07668, E = 0.0023, F = 1.07003 $.}
        \label{2}
    \end{minipage}
\end{figure}

\subsection{Energy conditions}

\hspace{0.5cm}We have used the energy conditions to confirm that our model is physically sound. The energy conditions characterize the essential physical properties common to a wide range of matter states and standard non-gravitational fields within the framework of modern physics. Furthermore, the energy conditions provide robust constraints that allows to eliminate various non-physical equations that do not represent plausible states of existence. A complete evaluation of the model's physical viability requires a validation against all the standard energy conditions to ensure that the matter distribution remains physically permissible. The standard energy conditions are mainly Null Energy Condition $(NEC)$, Weak Energy Condition $(WEC)$, Strong Energy Condition $(SEC)$ and Dominant Energy Condition $(DEC)$. The physical viability of the matter depends on the satisfaction of these energy conditions across the entire domain. The formulation for these conditions are expressed as follows:
\begin{equation}\label{24}
NEC: \rho + p \geq 0
\end{equation}
\begin{equation}\label{25}
WEC:\rho \geq 0, \hspace{0.5cm} \rho + p \geq 0
\end{equation}
\begin{equation}\label{26}
SEC:\rho + 3p \geq 0
\end{equation}
\begin{equation}\label{27}
DEC:\rho - p \geq 0.
\end{equation}\\
As illustrated in figures (\ref{3}-\ref{8}), we can see that our model remains physically sound. For different parameter values, we have ensured that every energy condition is satisfied consistently through the star's entire interior.

\begin{figure}[htbp]
    \centering
    \begin{minipage}{0.45\textwidth}
        \centering
        \includegraphics[width=\linewidth]{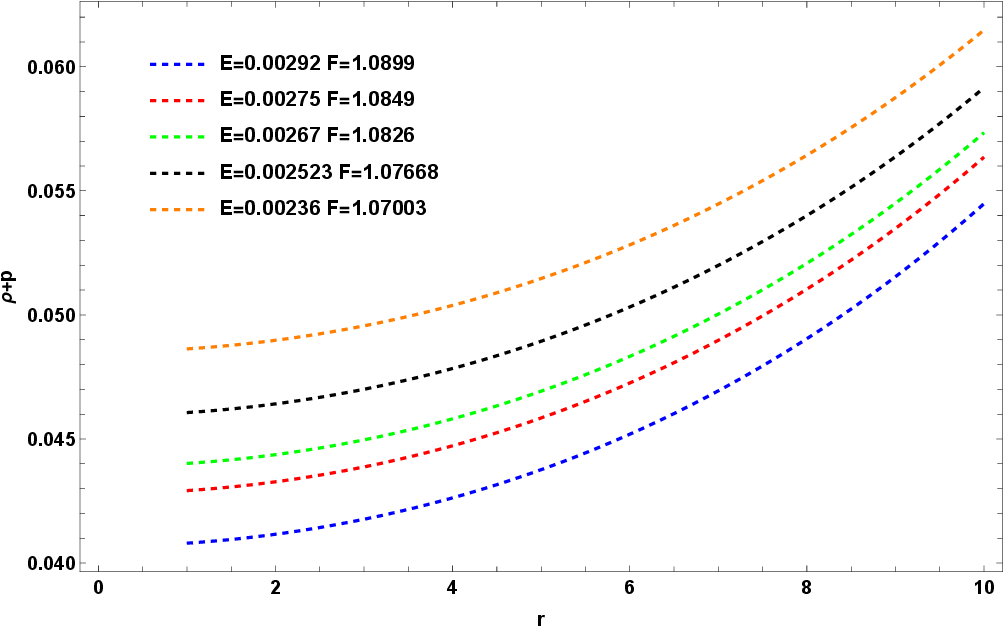}
        \caption{Graph of $ \rho + p $ w.r.t radial co-ordinate for CWS EoS.}
        \label{3}
    \end{minipage}
    \hfill
    \begin{minipage}{0.45\textwidth}
        \centering
        \includegraphics[width=\linewidth]{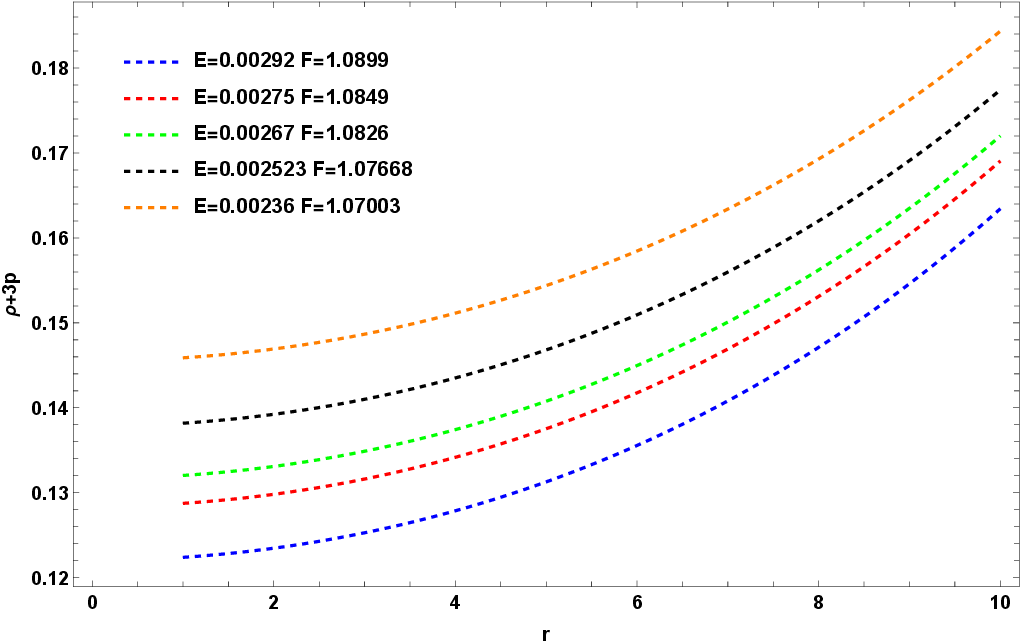}
        \caption{Graph of $ \rho + 3p $ w.r.t radial co-ordinate for CWS EoS.}
        \label{4}
    \end{minipage}
\end{figure}

\begin{figure}[htbp]
    \centering
    \begin{minipage}{0.45\textwidth}
        \centering
        \includegraphics[width=\linewidth]{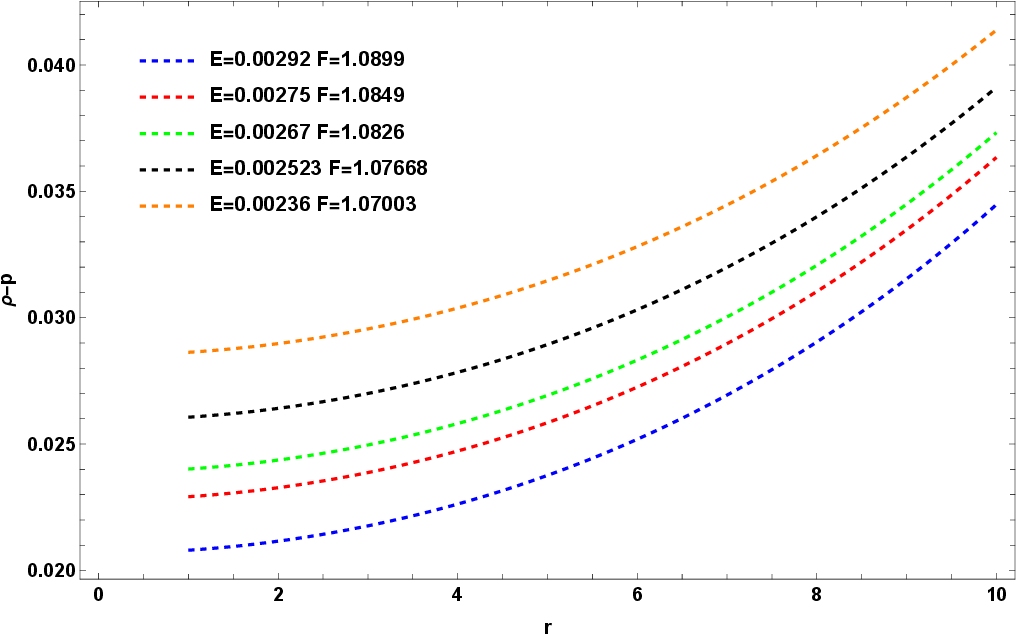}
        \caption{Graph of $ \rho - p $ w.r.t radial co-ordinate for CWS EoS.}
        \label{5}
    \end{minipage}
    \hfill
    \begin{minipage}{0.45\textwidth}
        \centering
        \includegraphics[width=\linewidth]{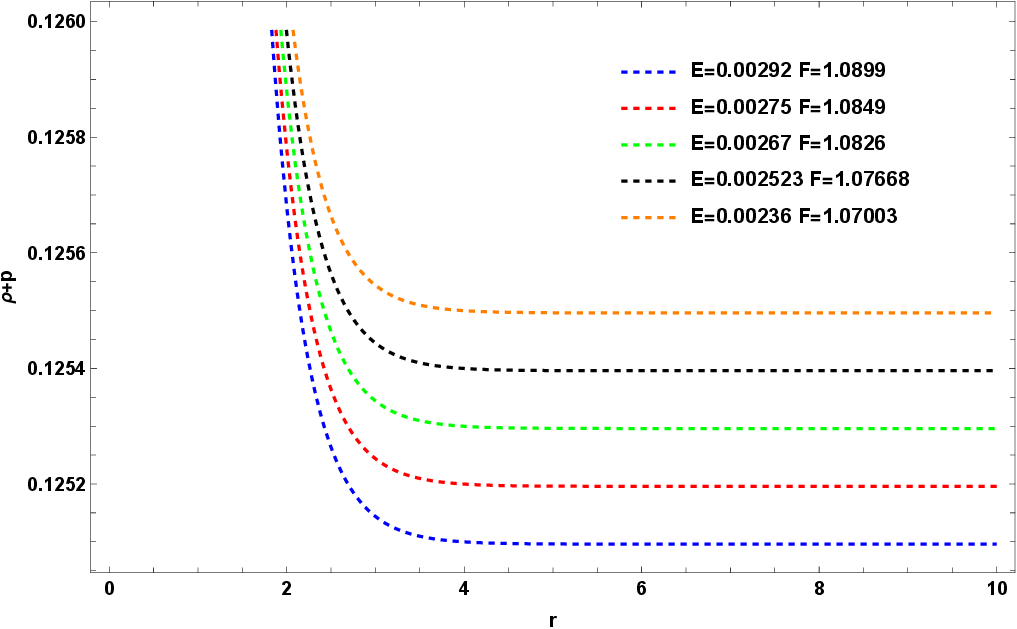}
        \caption{Graph of $ \rho + p $ w.r.t radial co-ordinate for GP EoS.}
        \label{6}
    \end{minipage}
\end{figure}

\begin{figure}[htbp]
    \centering
    \begin{minipage}{0.45\textwidth}
        \centering
        \includegraphics[width=\linewidth]{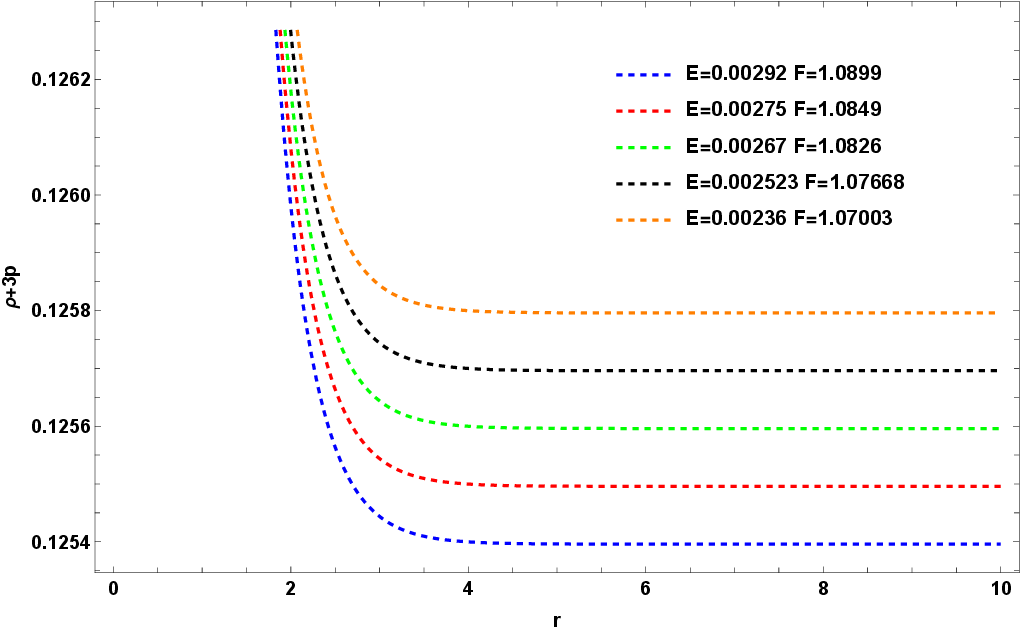}
        \caption{Graph of $ \rho + 3p $ w.r.t radial co-ordinate for GP EoS.}
        \label{7}
    \end{minipage}
    \hfill
    \begin{minipage}{0.45\textwidth}
        \centering
        \includegraphics[width=\linewidth]{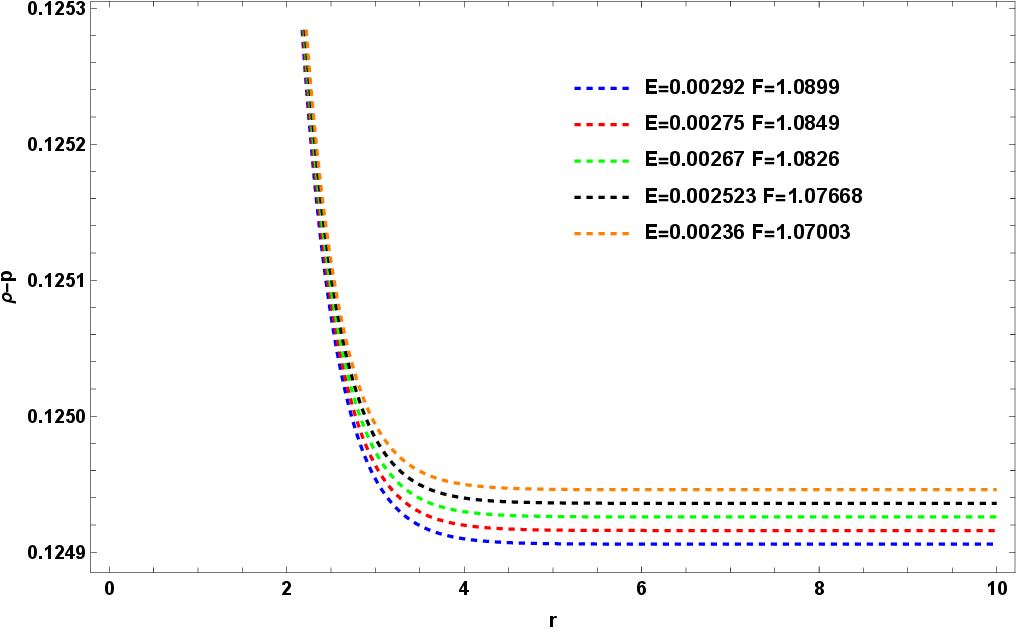}
        \caption{Graph of $ \rho - p $ w.r.t radial co-ordinate for GP EoS.}
        \label{8}
    \end{minipage}
\end{figure}

\clearpage

\subsection{Equation of state parameter}

\hspace{0.5cm}The EoS parameter for our model, which describes the relationship between the internal pressure and the energy density is defined by the following expression as\\
\begin{equation}\label{28}
\omega = \frac{p}{\rho}.
\end{equation}\\
Figures (\ref{9}) and (\ref{10}) illustrate the numerical evolution of the EoS parameter for the CWS and the GP cases respectively. It is evident from these plots that the parameter remains within the physically bounds of $ 0 < \omega < 1 $ and exhibits a monotonic decrease toward the stellar boundary. These results suggest that the medium behaves as a physically viable cosmological fluid devoid of exotic properties. The parameter range is bounded by two limits - the pressureless dust case $(\omega = 0)$ and the maximally stiff Zel'dovich fluid $(\omega = 1)$. Additionally, we have plotted the gradients for the energy-momentum components in the figures (\ref{11}-\ref{14}). From these plots it is clear that the gradients remain positive and follow a smooth steady pattern exactly as a stable star should in both the cases.

\begin{figure}[htbp]
    \centering
    \begin{minipage}{0.45\textwidth}
        \centering
        \includegraphics[width=\linewidth]{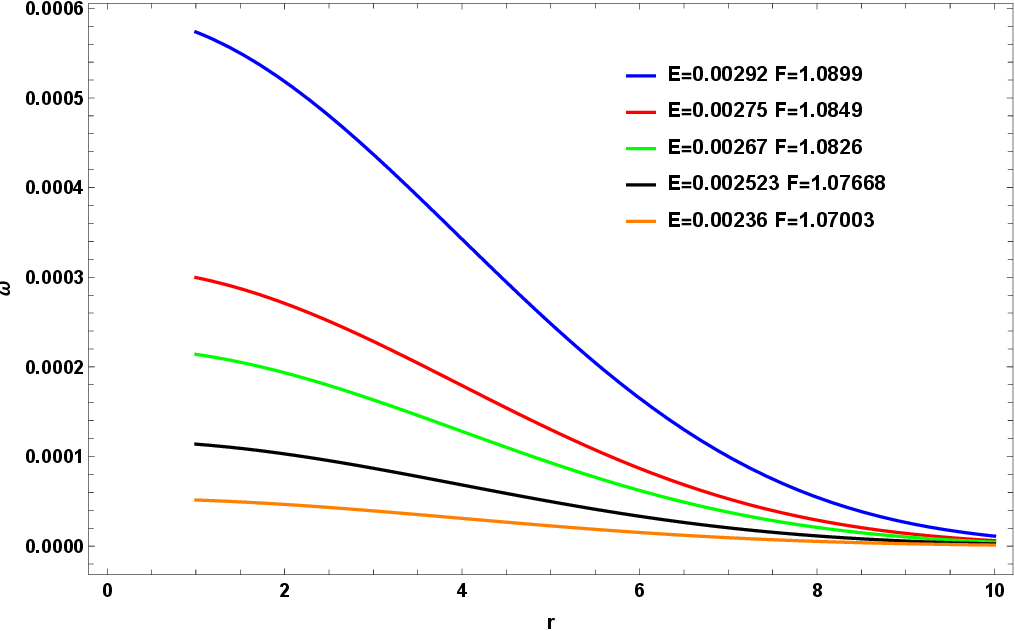}
        \caption{Variation of $ \omega $ for CWS EoS.}
        \label{9}
    \end{minipage}
    \hfill
    \begin{minipage}{0.45\textwidth}
        \centering
        \includegraphics[width=\linewidth]{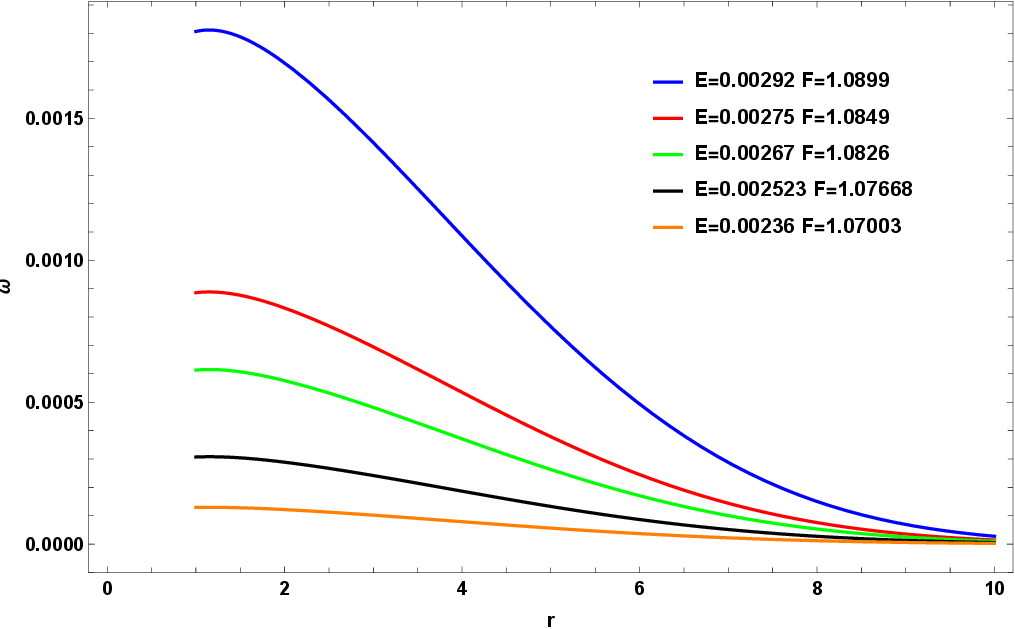}
        \caption{Variation of $ \omega $ for GP EoS.}
        \label{10}
    \end{minipage}
\end{figure}

\begin{figure}[htbp]
    \centering
    \begin{minipage}{0.45\textwidth}
        \centering
        \includegraphics[width=\linewidth]{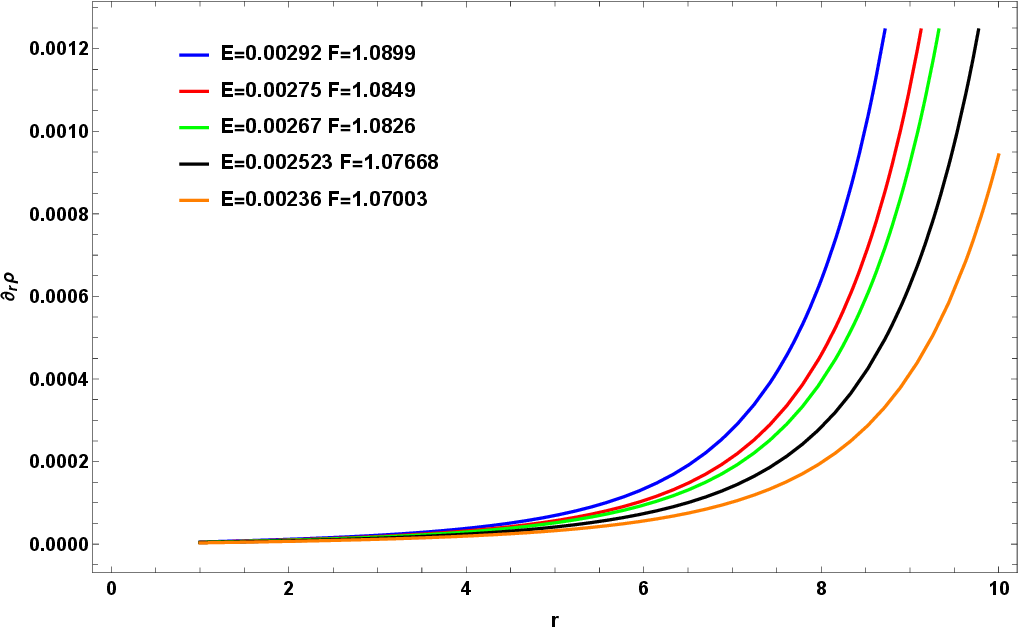}
        \caption{Figure of $ \partial_{r} \rho $ for CWS EoS.}
        \label{11}
    \end{minipage}
    \hfill
    \begin{minipage}{0.45\textwidth}
        \centering
        \includegraphics[width=\linewidth]{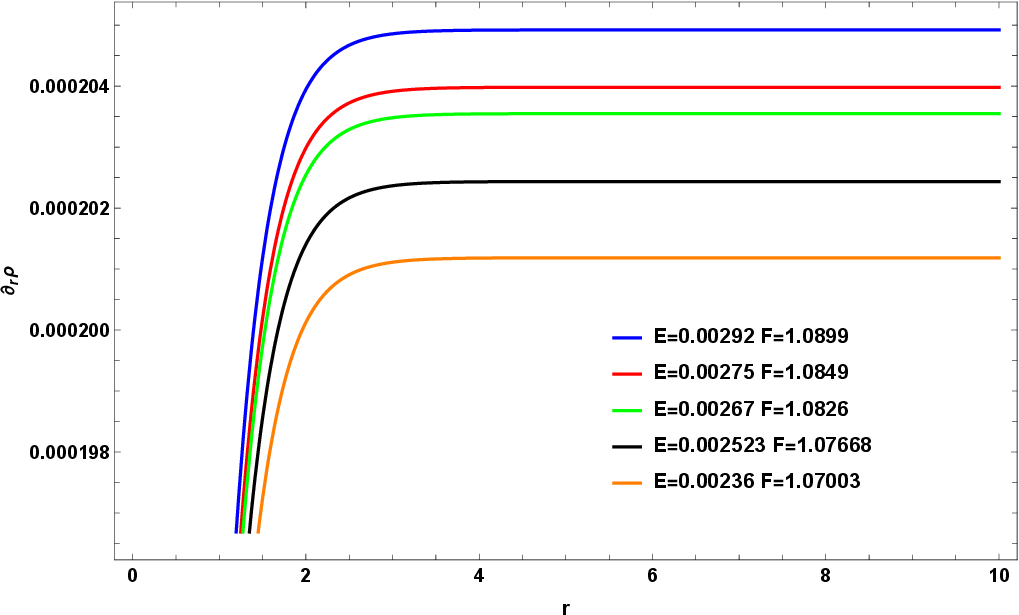}
        \caption{Figure of $ \partial_{r} \rho $ for GP EoS.}
        \label{12}
    \end{minipage}
\end{figure}

\begin{figure}[htbp]
    \centering
    \begin{minipage}{0.45\textwidth}
        \centering
        \includegraphics[width=\linewidth]{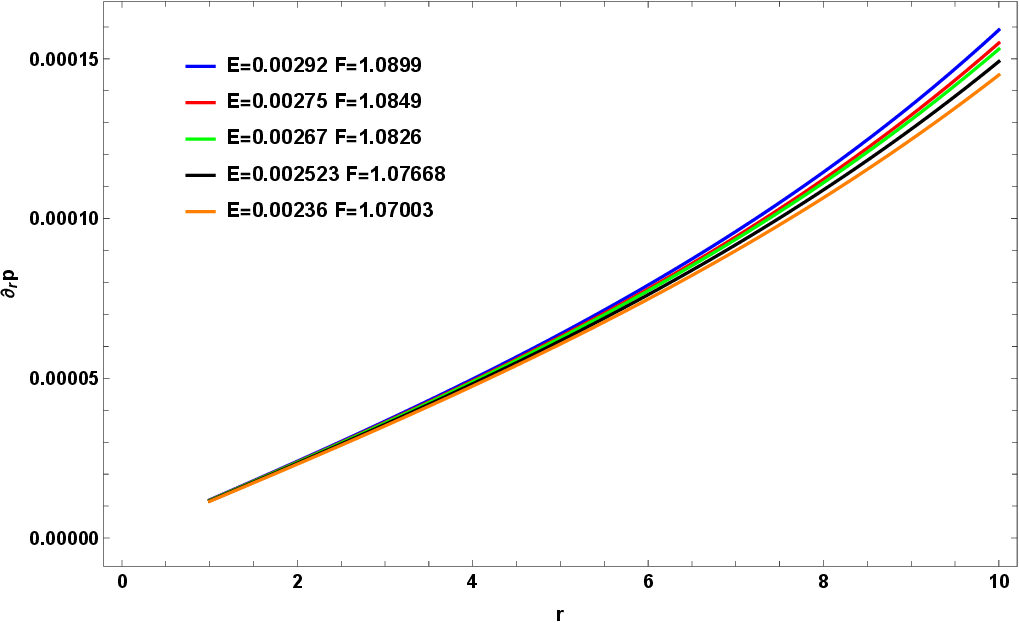}
        \caption{Figure of $ \partial_{r} p $ for CWS EoS.}
        \label{13}
    \end{minipage}
    \hfill
    \begin{minipage}{0.45\textwidth}
        \centering
        \includegraphics[width=\linewidth]{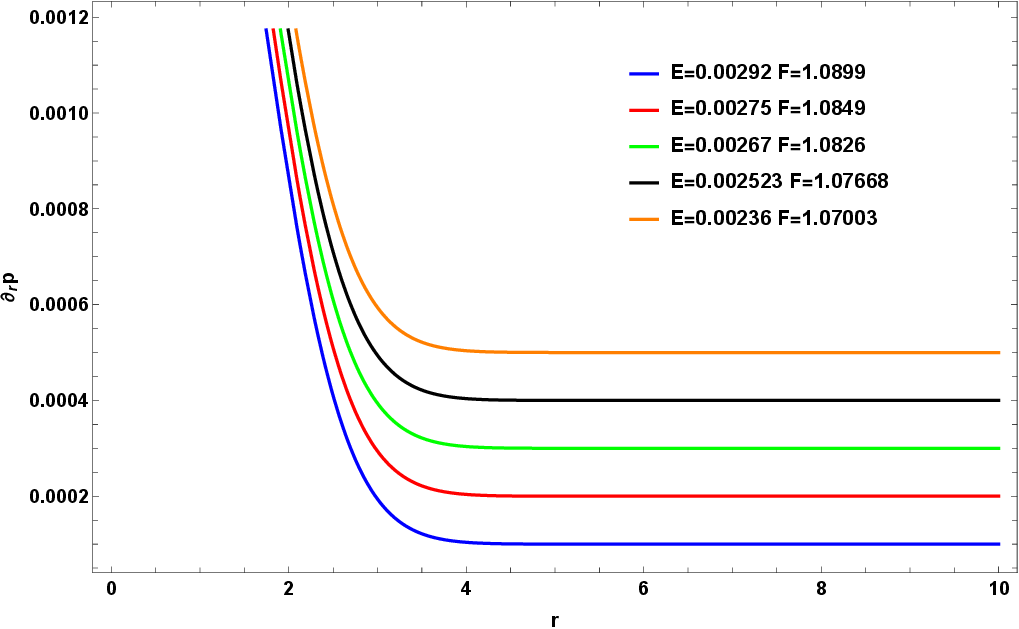}
        \caption{Figure of $ \partial_{r} p $ for GP EoS.}
        \label{14}
    \end{minipage}
\end{figure}

\section{Stability analysis}\label{sec4}

\hspace{1cm}Next, we have examined whether these BEC stars can actually remain stable under dRGT massive gravity. The stability of the equilibrium configuration is assessesed through the radial behavior of the redshift function, the calculation of the relativistic adiabatic index and the application of Herrera's cracking stability criterion.\\

\subsection{Redshift function}

\hspace{0.5cm}In astrophysics, study of redshift is very important as it predicts a way to study the characteristics of our galaxy and the broader Universe. This process is demonstrated by a displacement of electromagnetic emission towards the lower frequency, higher wavelength region of the spectral distribution. The redshift parameter represents the fractional shift in the wavelength of electromagnetic radiation between emission and detection.  Photons originating from the stellar center experience significant dispersion and gravitational energy loss as they traverse the high density core region. It is given by \cite{d1}
\begin{equation}\label{29}
{\texttt{Z}}_{sur} = \frac{1}{\sqrt{g_{tt}}} - 1.
\end{equation}\\
The surface redshift of a relativistic fluid configuration is constrained by a maximum theoretical value of $ {\texttt{Z}}_{sur} < 5.211 $ \cite{d2}. Figures (\ref{15}) and (\ref{16}) confirm that the surface redshift values are strictly positive for our model and also it remain below the threshold of $ 2 $, ensuring that the model adheres to standard relativistic limits for compact objects. The observed redshift profile thus validates the construction for both the CWS and GP EoS models. Consequently, the stellar structure satisfies the fundamental requirements for physical consistency across the entire manifold.

\begin{figure}[htbp]
    \centering
    \begin{minipage}{0.45\textwidth}
        \centering
        \includegraphics[width=\linewidth]{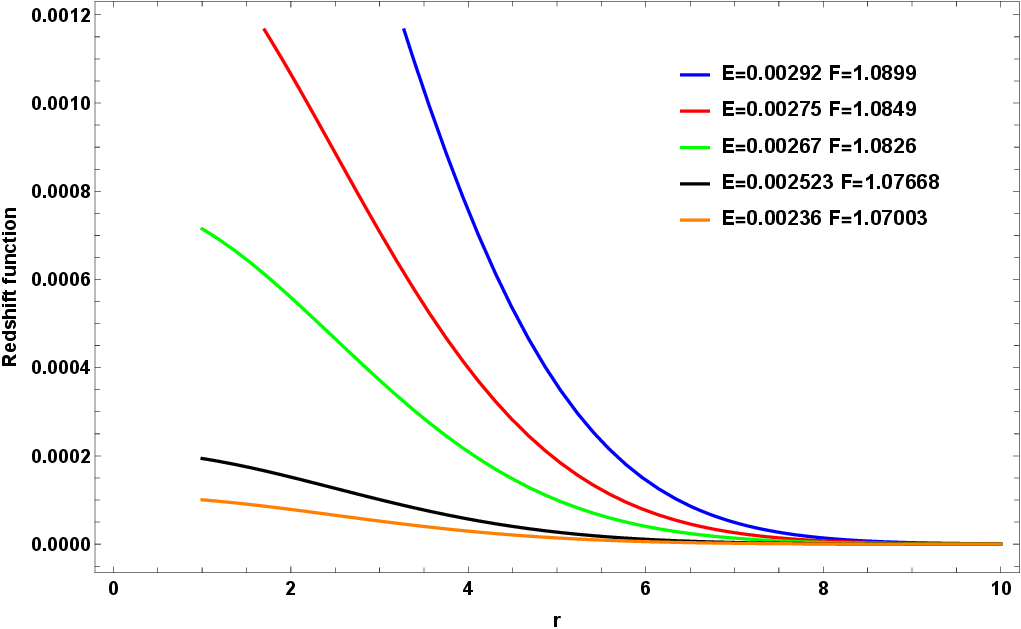}
        \caption{Variation of $ {\texttt{Z}}_{sur} $ for CWS profile.}
        \label{15}
    \end{minipage}
    \hfill
    \begin{minipage}{0.45\textwidth}
        \centering
        \includegraphics[width=\linewidth]{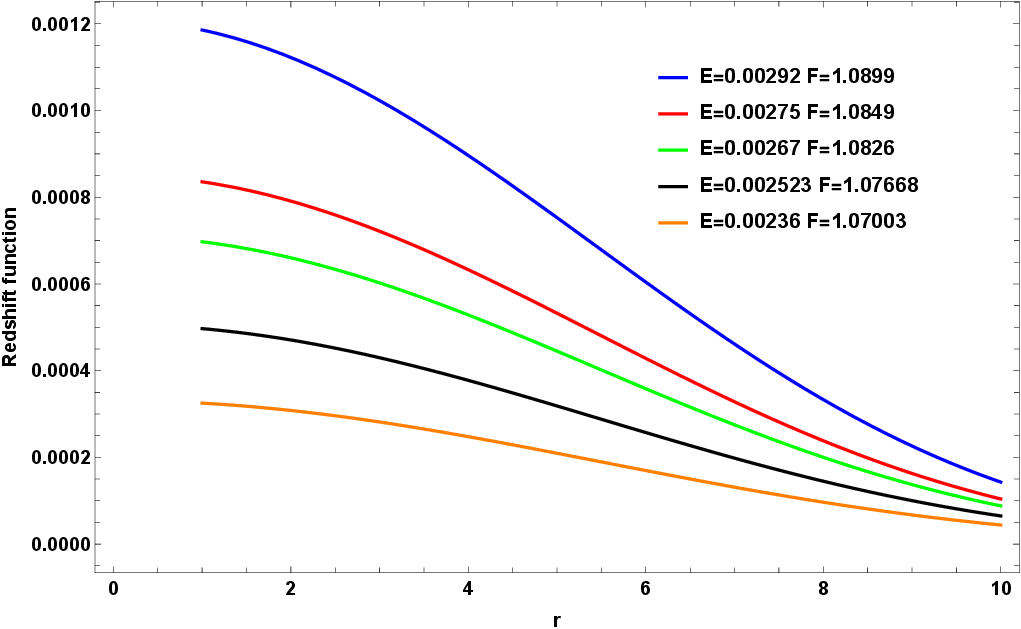}
        \caption{Variation of $ {\texttt{Z}}_{sur} $ for GP profile.}
        \label{16}
    \end{minipage}
\end{figure}

\subsection{Adiabatic index}

\hspace{0.5cm}The relativistic adiabatic index $(\Gamma)$, which is defined as the ratio of specific heats provides a fundamental tool for evaluating the dynamical stability of the proposed stellar configuration. We have the formulation for $(\Gamma)$ as \cite{d3}\\
\begin{equation}\label{30}
\Gamma = \big(\frac{\rho + p}{p}\big)\frac{dp}{d\rho}.
\end{equation}\\
To ensure the viability, the adiabatic index must exceed the critical value of $ \frac{4}{3} $ throughout the interior, as established in the works of \cite{d4}. The radial evolution of the adiabatic index within the dRGT massive gravity framework for our study is illustrated in figures (\ref{17}) and (\ref{18}). We can see that these profiles exhibit a more or less monotonic nature for various parameter values, but consistently exceeding the critical value of $ \frac{4}{3} $ throughout the stellar interior. Thus, the validation from these plots confirm the dynamical stability of our model from a relativistic perspective. 

\begin{figure}[htbp]
    \centering
    \begin{minipage}{0.45\textwidth}
        \centering
        \includegraphics[width=\linewidth]{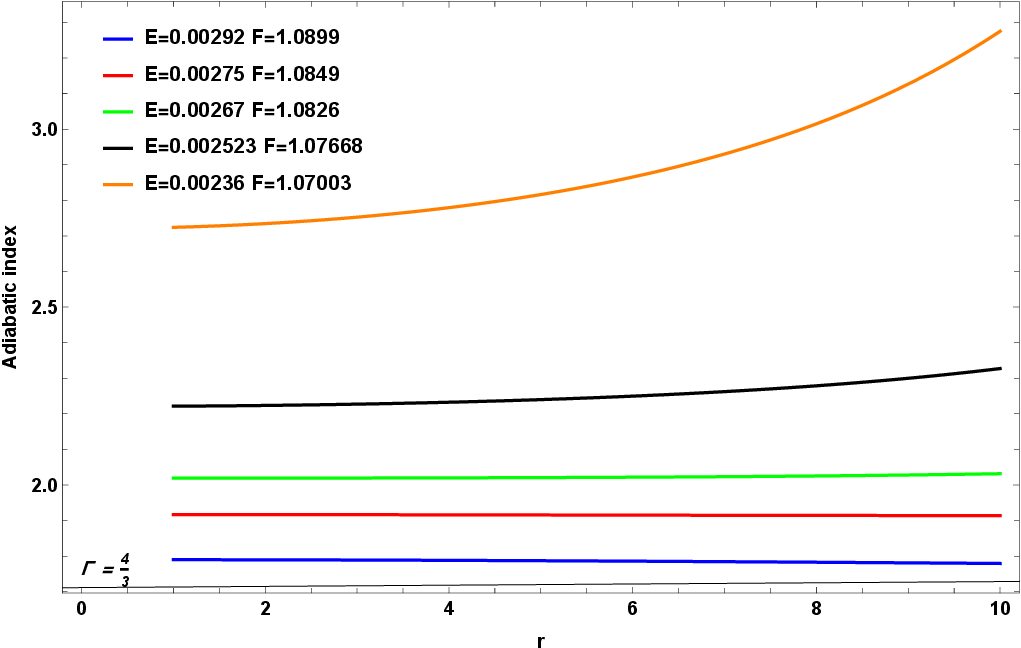}
        \caption{Nature of $ \Gamma $ for CWS EoS.}
        \label{17}
    \end{minipage}
    \hfill
    \begin{minipage}{0.45\textwidth}
        \centering
        \includegraphics[width=\linewidth]{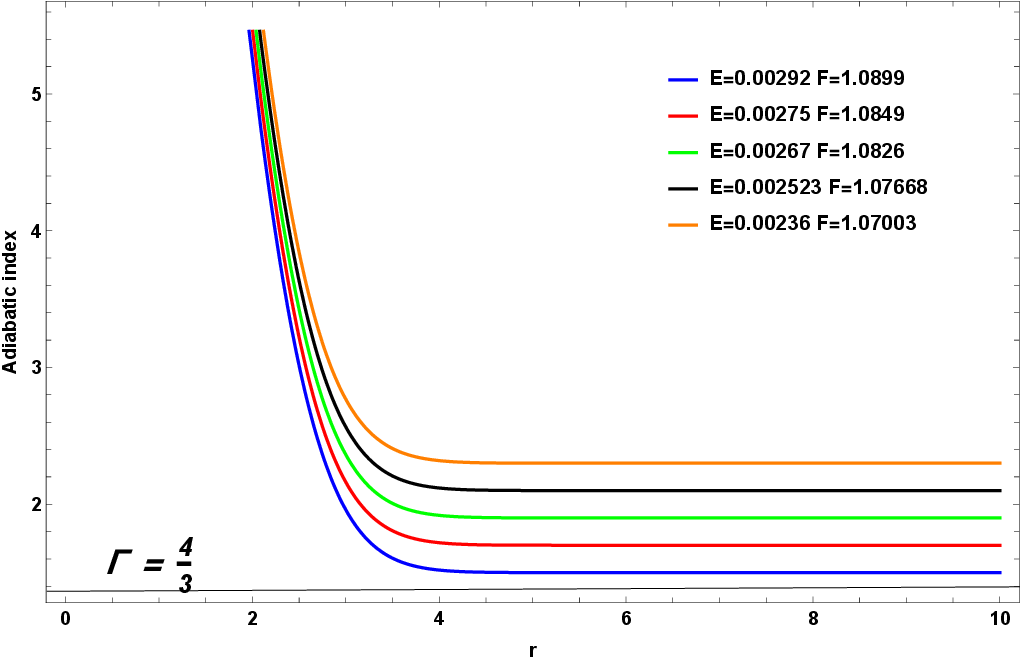}
        \caption{Nature of $ \Gamma $ for GP EoS.}
        \label{18}
    \end{minipage}
\end{figure}

\subsection{Herrera's condition}

\hspace{0.5cm}The stability of our proposed model is further validated using Herrera's casuality criteria, which is highly essential in the study of relativistic compact objects \cite{TC1,TC2}. To maintain casuality, the square of the sound speed $ V^{2} $ is required to satisfy the constraint $ 0 \leq V^{2} \leq 1 $. The parameter is given as\\
\begin{equation}\label{31}
V^{2} = \frac{dp}{d\rho} \leq c^{2} = 1.
\end{equation}\\
Figures (\ref{19}) and (\ref{20}) below depict the radial variation of the sound speed square, which illustrated how this velocity profile evolves from the stellar core towards the surface for our case. These figures clearly indicate that the sound speed square $ V^{2} $ remain confined within the casuality bounds of $ (0,1) $ across the entire stellar system for both the EoSs. While a monotonic decreasing nature is observed, the values consistently remain positive and also  sub-luminal, thereby reinforcing the physical consistency and viability of the proposed configuration.

\begin{figure}[htbp]
    \centering
    \begin{minipage}{0.45\textwidth}
        \centering
        \includegraphics[width=\linewidth]{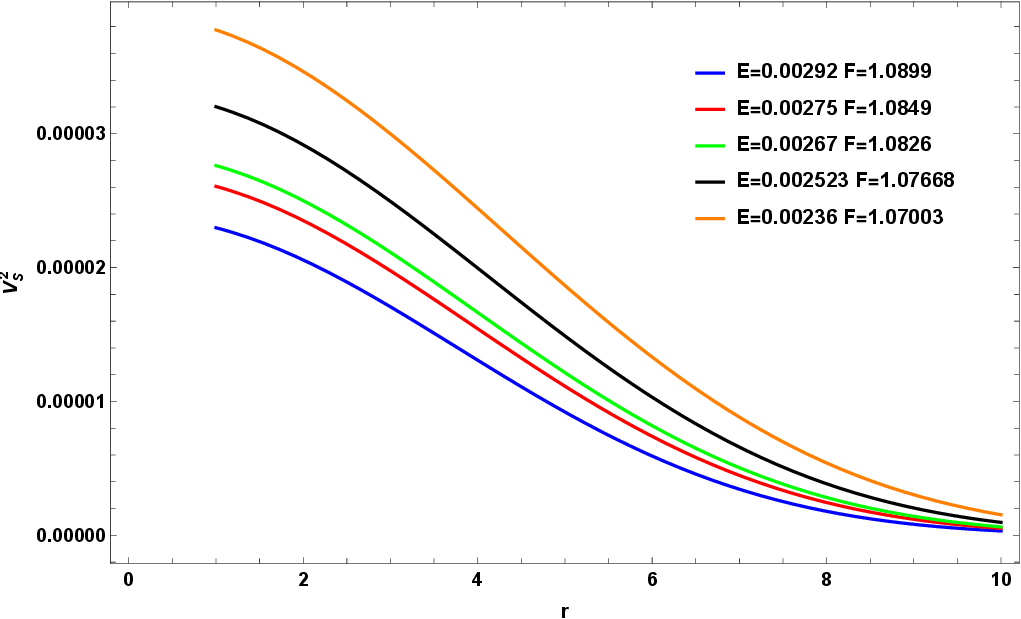}
        \caption{Plot of $ V^{2} $ for CWS EoS.}
        \label{19}
    \end{minipage}
    \hfill
    \begin{minipage}{0.45\textwidth}
        \centering
        \includegraphics[width=\linewidth]{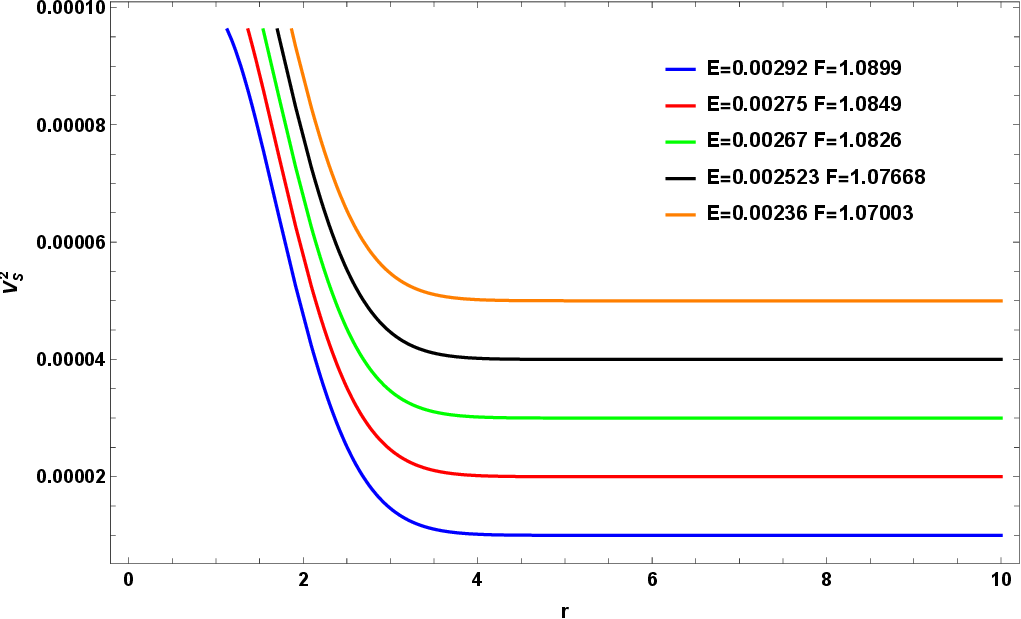}
        \caption{Plot of $ V^{2} $ for GP EoS.}
        \label{20}
    \end{minipage}
\end{figure}

\section{Discussions and conclusion}\label{sec5}

\hspace{1cm}In this paper, we have explored the features of BECs and how they behave within the core of compact stellar objects. This research offers an evaluation of spherically static and symmetric configurations under dRGT massive gravity, utilizing the Kuchowicz approach to derive the underlying metric solutions. Numerical solutions were derived by integrating the governing field  equations with CWS and GP EoS for the BEC stellar matter. The gravitational field equations were resolved through numerical integration to produce the stellar model, with the internal solutions matched smoothly to the exterior vacuum in accordance to the required boundary conditions. To examine the influence of the Kuchowicz parameter under this massive gravity framework, we have utilized the specific stellar candidates $ PSR JI6142230, Vela X-1, PSR J1903+327, Cen X-3 $ and $ SMC X-1 $ as benchmarks. The resulting data proved to be physically consistent and compliant with standard models that are well supported by current astrophysical configurations.\\
The acceptability of the proposed stellar model is reinforced by the fulfillment of the fundamental energy conditions as illustrated in figures (\ref{3}-\ref{8}). Additionally, the evolution of the EoS parameters in figures (\ref{9}-\ref{10}) and the gradients of the energy momentum tensor components in figures (\ref{11}-\ref{14}) demonstrate that the matter distribution remains physically viable and consistent with the relativistic requirements.\\
The radial evolution and behavior of the gravitational redshift function for both CWS and GP EoSs are graphically illustrated in figures (\ref{15}) and (\ref{16}). The figures illustrate that the physical profiles exhibit continuity under both the cases. Furthermore, these values remain within the permissible bounds $0$ and $2$, which confirms the viability of our proposed model. The distribution of the radial profile of the adiabatic index is depicted in figures (\ref{17}) and (\ref{18}), which exhibits a monotonic nature outwards from the core. Notably, the profile remains regular and finite at the stellar center, satisfying the conditions for a non-singular interior. In the context of spherically symmetric perfect fluid distribution, within this dRGT massive gravity the adiabatic index consistently exceeds the critical value of $ \frac{4}{3} $ threshold, as we can see from the figures which validates the dynamic stability of the proposed stellar configuration. In accordance with the Herrera's stability criteria, the squared sound speed component must satisfy the inequality $ 0 \leq V^{2} \leq 1 $. In our work, numerical solutions from  figures (\ref{19}) and (\ref{20}) show that this casuality condition is maintained throughout the stellar interior, ensuring the physical consistency of the model. While it drops steadily towards the surface, the value never falls below $ 0 $, which helps to keep the overall structure stable.\\
We can thus conclude that we have carried out a detailed investigation of the BEC stars within the dRGT massive gravity framework, adopting the Kuhowicz metric in conjunction with the CWS and GP EoS. Our findings confirm that the constructed stellar models are physically viable, regular and consistent. These results establish the present framework as a reliable platform for probing the properties of compact objects governed by BECs under different modified gravities under different physical scenarios in the near future.\\

\textbf{Data Availability Statement}: The manuscript has no associated data.\\
\textbf{Funding Acknowledgement}: No financial support was received for this study.\\

\end{document}